\documentclass[a4paper]{article}

\usepackage{icrc2011}

\title{Exploring the Galaxy at TeV energies: Latest results from the H.E.S.S.~Galactic Plane Survey.}

\shorttitle{H.~Gast for the H.E.S.S.~Collaboration}

\authors{H.~Gast$^{1}$, F.~Brun$^{2}$, S.~Carrigan$^{1}$, R.C.G.~Chaves$^{1}$, C.~Deil$^{1}$, A.~Djannati-Ata\"{i}$^{3}$, Y.~Gallant$^{4}$, V.~Marandon$^{3}$, M.~de Naurois$^{2}$, and R.~de los Reyes$^{1}$, for the H.E.S.S.~collaboration}
\afiliations{$^1$Max-Planck-Institut f\"ur Kernphysik, Saupfercheckweg 1, 69117 Heidelberg, Germany\\
$^2$Laboratoire Leprince-Ringuet, Ecole Polytechnique, CNRS/IN2P3, F-91128 Palaiseau, France\\
$^3$Astroparticule et Cosmologie (APC), CNRS, Universit\'{e} Paris 7 Denis Diderot, 10, rue Alice Domon et L\'{e}onie Duquet, F-75205 Paris Cedex 13, France\\
$^4$Laboratoire Univers et Particules de Montpellier, Universit\'e Montpellier 2, CNRS/IN2P3,  CC 72, Place Eug\`ene Bataillon, F-34095 Montpellier Cedex 5, France}
\email{henning.gast@mpi-hd.mpg.de}

\abstract{The High Energy Stereoscopic System (H.E.S.S.) is an array
of four imaging atmospheric-Cherenkov telescopes located in Namibia
and designed to detect extensive air showers initiated by $\gamma$-rays
in the very-high-energy domain. It is an ideal
instrument for surveying the Galactic plane in search of new sources,
thanks to its location in the Southern Hemisphere, its
excellent sensitivity, and its large field-of-view. The efforts of the
H.E.S.S. Galactic Plane Survey, the first comprehensive survey of the
inner Galaxy at TeV energies, have contributed to the discovery of an
unexpectedly large and diverse population of over 60 sources of VHE
$\gamma$-rays within its current range of $\ell=250$ to 65 degrees in
longitude and $|b|\leq{}3.5$ degrees in latitude. The population of VHE
$\gamma$-ray emitters is dominated by the pulsar wind nebula and
supernova remnant source classes, although nearly a third remain
unidentified or confused.

The sensitivity of H.E.S.S. to sources in the inner Galaxy has
improved significantly over the past two years, from continued survey
observations, dedicated follow-up observations of interesting source
candidates, and from the development of advanced methods for
discrimination of $\gamma$-ray-induced showers from the dominant
background of hadron-induced showers. The latest maps of the Galaxy at
TeV energies will be presented, and a few remarkable new sources will
be highlighted.}
\keywords{H.E.S.S., Galactic Plane Survey, gamma-ray astronomy, supernova remnant, pulsar wind nebula}

\begin{document}
\maketitle

\section{Introduction}
At the far end of the energy spectrum accessible to astronomical
observations, imaging atmospheric Cherenkov telescopes (IACTs) are
employed to detect extended air showers initiated by very-high-energy
(VHE; $E>0.1\,\mathrm{TeV}$) photons.
With the latest generation of IACTs, the detection of
VHE $\gamma$-rays has turned into a mature astronomical discipline,
and more than 120 sources are currently listed in the online TeV
$\gamma$-ray catalogue TeVCat~\footnote{{\tt http://tevcat.uchicago.edu/}}. Here we report on the
status and latest results of the Galactic Plane Survey (GPS),
undertaken with the High Energy Stereoscopic System (H.E.S.S.). With a sizable fraction of the annual
observation budget of H.E.S.S.~dedicated to it, the GPS aims at
performing a systematic scan of the inner Galaxy, with the main goal
of discovering previously unknown emitters of VHE $\gamma$-rays.
More than 60 Galactic VHE $\gamma$-ray sources are now known. The
population is dominated by sources that are linked to the final stages
in stellar evolution, namely pulsar wind nebulae (PWNe) and
supernova remnants (SNRs). For nearly a third of the sources, however,
no plausible counterpart at other wavelengths has been found yet, or
the physical origin of the detected emission remains unclear.\\
\par
H.E.S.S. consists of four identical $12\,\mathrm{m}$ diameter IACTs located at an
altitude of $1800\,\mathrm{m}$ above sea level in the Khomas Highlands
of Namibia~\cite{ref:crab}. Its location in the Southern Hemisphere
affords it an excellent view of the inner Galaxy with low energy threshold. Each of
the four H.E.S.S.~telescopes is equipped with a camera containing 960
photomultiplier tubes and a tesselated mirror with a combined area of
$107\,\mathrm{m}^2$~\cite{ref:konrad03}. The optical design allows for a comparatively large, $5^\circ$
field-of-view (FoV). The H.E.S.S. array has an angular resolution of $\sim\!0.1^\circ$ and
an energy resolution of $\sim\!15\,\%$.
Its high sensitivity, coupled with the large FoV, permits
H.E.S.S.~to effectively survey large areas of the Galaxy
within a reasonable amount of time.

\section{The H.E.S.S.~Galactic Plane Survey}
\begin{figure*}[p]
\centering
\includegraphics[height=21cm]{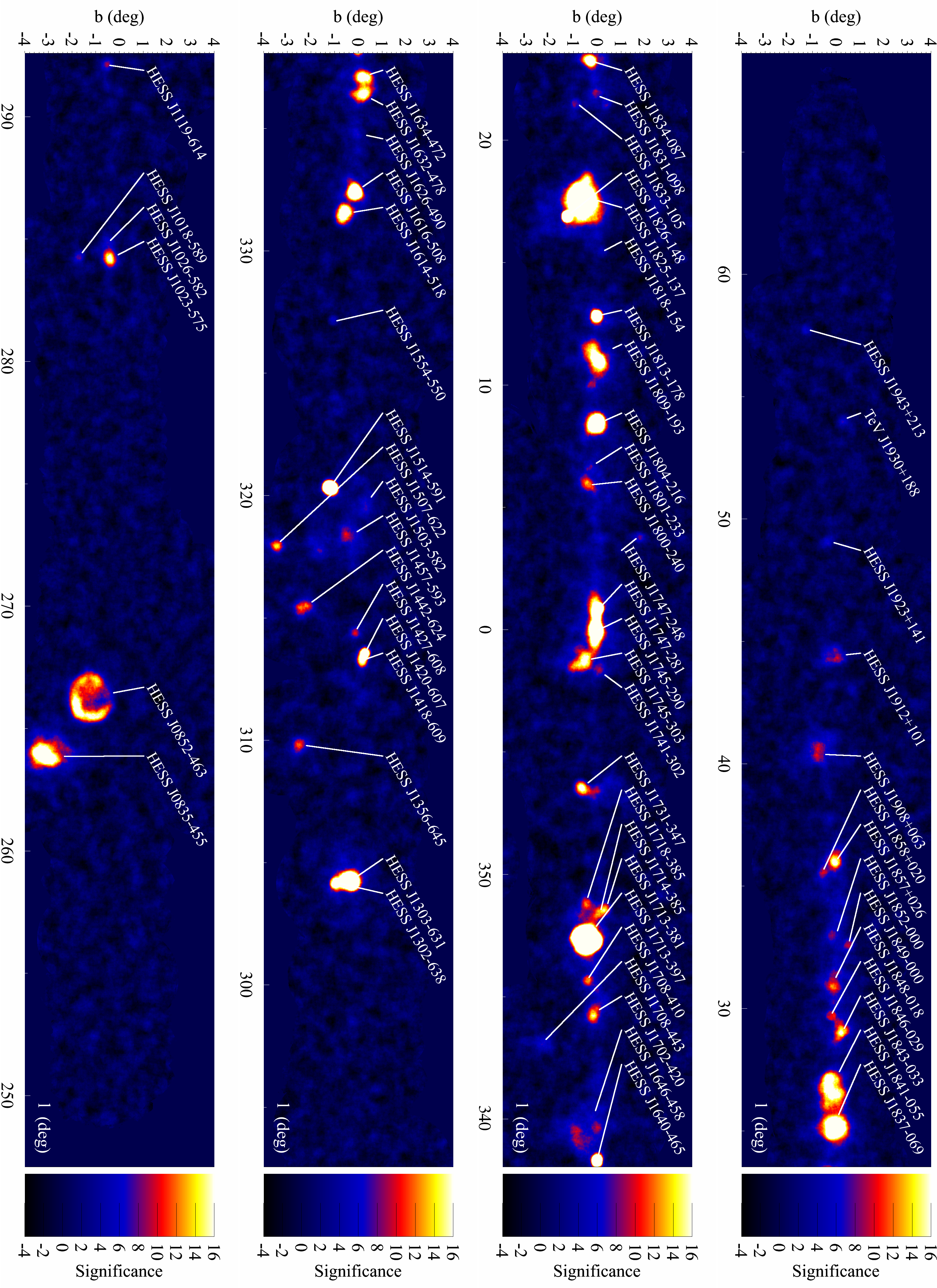}
\caption{
Latest significance map for the H.E.S.S.~Galactic Plane Survey. The
pre-trials significance for a correlation radius of $0.22^\circ$ is shown. The
colour transition from blue to red corresponds to $\sim\!5\sigma$ post-trials
significance. The trial factor takes into account the fact that many
sky positions are tested for an excess above the background, thus
increasing the chance of finding a random upward fluctuation of the
background. The map has been filled for regions on the sky where the
sensitivity of H.E.S.S.~for point sources ($5\sigma$ pre-trials, and assuming
the spectral shape of a power law with index 2.5) is better than
10$\,\%$ Crab.}
\label{fig:surveysig}
\end{figure*}

\begin{figure*}[t]
\centering
\includegraphics[width=16cm]{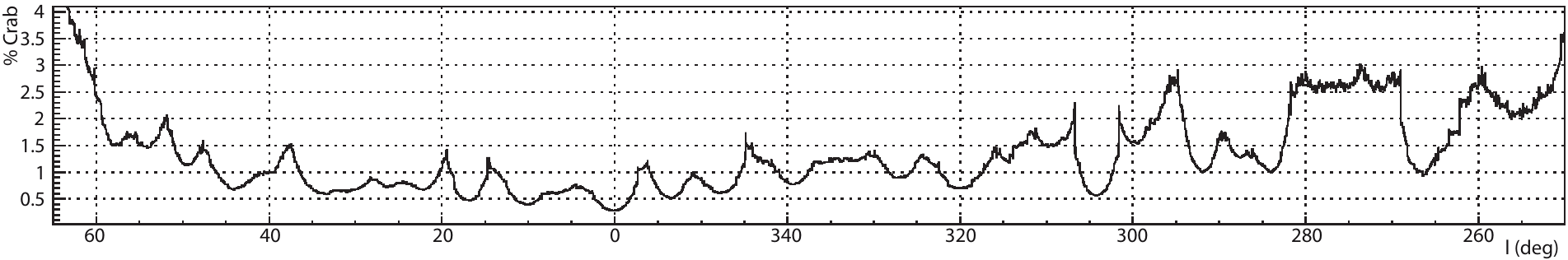}
\caption{Sensitivity of H.E.S.S.~to point-like $\gamma$-ray sources
with an assumed spectral index of $2.5$, for a detection level of
$5\sigma$ pre-trial, at $b=-0.3^\circ$, the approximate
average latitude of Galactic sources. The sensitivity is expressed in
units of the Crab integral flux
$F(\geq{}1\,\mathrm{TeV})=2.26\,\cdot\,10^{-7}\,\mathrm{m}^{-2}\,\mathrm{s}^{-1}$.}
\label{fig:sensitivity}
\end{figure*}

Most of the available observation time of H.E.S.S.~is
spent looking at pre-defined targets that seem promising because of
their known astrophysical properties. In the H.E.S.S.~Galactic Plane
Survey, a different approach is followed. Here, the inner Galaxy is
systematically scanned using observation positions with overlapping
fields-of-view, with the main goal of discovering new
$\gamma$-ray sources and enabling population studies of
Galactic source classes as a consequence. Advanced analysis techniques for background
suppression, e.g.~\cite{ref:tmva,ref:model,ref:opttmva,ref:apcmva},
play a very important role in this endeavour.

Over the course of its operation, H.E.S.S.~has assembled an impressive
dataset in the region of the inner Galaxy.
In the last major report on the H.E.S.S.~Galactic Plane Survey in January
2006~\cite{ref:survey}, the detection of 17 sources in the range $\ell\,\pm\,30^\circ$ and
$b\,\pm\,3^\circ$ was presented using a dataset comprising $230\,\mathrm{h}$ of
observation time, corrected for readout dead-time.
Since then, this dataset has increased dramatically; it now includes
over $2300\,\mathrm{h}$ of data covering the longitude range from
$\ell=250^\circ$ to $65^\circ$, and more than 60
Galactic sources have been detected as a result.
The total dataset that constitutes the Galactic Plane Survey is made
up of observations taken with two different strategies: Firstly, in {\it
scan mode}, pointings are distributed systematically along the
Galactic plane, usually in three strips in latitude ($b=-1,0,+1^\circ$) and
with spacings of $\sim\!{}0.7^\circ$ in longitude. Deeper, {\it
pointed observations} are taken on promising source candidates,
and all other observations that fall within the survey region are included in the
dataset, too. For the past and ongoing observation campaigns in 2010
and 2011, the focus of the survey effort has been put on achieving a more
uniform exposure in the survey core region ($\ell=282^\circ$ to
$60^\circ$), and on deepening the exposure in the region from $\ell=268^\circ$ to
$282^\circ$.

Figure~\ref{fig:surveysig} shows the latest significance map obtained
for the survey region. After calibration and quality selection, a
multivariate analysis technique~\cite{ref:tmva} based on shower and image shape
parameters is used to discriminate $\gamma$-ray events from
cosmic ray-induced showers. A minimum image amplitude of
$160\,\mathrm{p.e.}$ is required. The remaining background is estimated by
the ring background technique~\cite{ref:crab}. Regions on the sky
containing known VHE $\gamma$-ray sources are excluded from the ring,
and the ring radius is adaptively enlarged where a large fraction of
the ring area overlaps with an excluded region. The significance value for each
position is then calculated~\cite{ref:lima}, by summing the candidate events within a
fixed and pre-defined correlation radius of $0.22^\circ$, suitable for
extended sources, and comparing to the estimated background level at
that position. The map now covers the full longitude range of the GPS,
and in particular, the extensions above $\ell=60^\circ$ and below
$\ell=275^\circ$, that are presented here for the first time.

Figure~\ref{fig:sensitivity} depicts the current sensitivity to
$\gamma$-ray sources, as an example for point-like sources emitting a
simple power-law spectrum with index
$\Gamma=2.5$ and located at a Galactic latitude of $b=-0.3^\circ$,
the approximate average among known Galactic sources. The sensitivity
is at the few-permille Crab level for the deepest exposures and is
below $2\,\%$ Crab for practically all of the longitude range $\ell=283^\circ$ to
$59^\circ$ at this latitude. Note that in
crowded regions, especially close to the Galactic plane,
the detection of new faint sources is
complicated by the foreground emission from stronger sources. The
official H.E.S.S.~source catalogue containing all sources published
in refereed journals is available online at {\tt www.mpi-hd.mpg.de/hfm/HESS/}.

\section{New discoveries}
In this section, we briefly spotlight some of the recent additions to
the list of Galactic H.E.S.S.~sources, many of which are presented at
this conference for the first time.\\
\par
{\bf Discovery of VHE emission towards the direction of SNR G284.3-1.8}\\
After $40\,\mathrm{h}$ of observations, H.E.S.S.~has detected
significant emission (HESS J1018-589) from the direction of SNR G284.3-1.8, an
incomplete radio shell with nonthermal spectrum and interacting with
molecular clouds~\cite{ref:g284}. An association with the Vela-like pulsar PSR
J1016-5857 can be made if distance measurements of 
the SNR and the pulsar as well as the large offset between the centre
of the shell and the  pulsar can be reconciled.\\
\par
{\bf Detection of High and Very High Energy $\gamma$-rays from the
direction of SNR~G318.2+0.1}\\
An extended source of VHE $\gamma$-rays (HESS J1457-593) was recently
discovered~\cite{ref:g318} in the direction towards the
shell-type SNR G318.2+0.1, with a hard spectrum
extending beyond $20\,\mathrm{TeV}$. The VHE $\gamma$-ray emission overlaps the southern
rim of the SNR and extends roughly $0.3^\circ$ outward from the shell. In the most likely scenario, cosmic
rays accelerated in the shock-front of the SNR interact with the
target material in a spatially coincident giant molecular cloud seen in
$^{12}\mathrm{CO}$ line data and produce $\gamma$-rays from
pion decay. This scenario is supported by public Fermi-LAT data
revealing $\gamma$-ray emission at the position of the H.E.S.S.~source
and extending towards the western rim of the SNR.\\
\par
{\bf A newly discovered VHE $\gamma$-ray PWN candidate around PSR J1459-6053}\\
Survey observations have revealed a significant VHE $\gamma$-ray
excess from the direction of PSR J1459--6053, a rather old
$\gamma$-ray pulsar (64 kyr) with a spindown energy of
$9\,\cdot\,10^{-35}\mathrm{erg}/\mathrm{s}$,
discovered by Fermi-LAT in high-energy $\gamma$-rays~\cite{ref:j1459}. The
X-ray pulsar counterpart has been recently detected using the Suzaku
satellite. The source is located in a region of the sky highly
populated with VHE sources, with at least two other confirmed VHE
$\gamma$-ray sources (PWN and SNR) within less than 2 degrees.\\
\par
{\bf Detection of VHE $\gamma$-ray emission from the intriguing
composite SNR G327.1-1.1}\\
SNR G327.1-–1.1 belongs to the category of SNRs
hosting a PWN (known as composite SNRs)
and exhibits a shell and a bright central PWN, both seen in radio and
X-rays. Interestingly, radio observations of the PWN show an
extended blob of emission and a curious narrow finger structure
pointing towards the offset compact X-ray source (the pulsar
candidate) indicating a possible fast moving pulsar in the
SNR or an asymmetric passage of the reverse shock.
Observations of the SNR G327.1-–1.1 with the H.E.S.S.~telescope
array resulted in the detection
of significant TeV $\gamma$-ray emission in spatial
coincidence with the PWN (HESS J1554-550)~\cite{ref:g327}.\\
\par
{\bf VHE $\gamma$-ray emission from the direction of Terzan~5}\\
H.E.S.S.~has discovered a new VHE $\gamma$-ray source (HESS J1747-248),
located in the immidiate vicinity of the Galactic globular cluster
Terzan~5~\cite{ref:t5}. The source appears extended and off-set from the cluster core
but overlaps significantly with Terzan~5. A random coincidence with the
globular cluster is unlikely ($\sim{}10^{-4}$) but this possibilty cannot firmly be
excluded. With the largest population of identified millisecond pulsars, a
very large core stellar density and the brightest GeV-range flux as
measured by Fermi-LAT, Terzan~5 stands out among Galactic globular
clusters. Interpretation of the available data accommodates several
possible origins for this VHE $\gamma$-ray source.\\
\par
{\bf Discovery of VHE $\gamma$-ray emission from the shell-type SNR
G15.4+0.1 with H.E.S.S.}\\
Statistically significant emission of VHE $\gamma$-rays (HESS
J1818-154) has now
been detected from the direction of the shell-type
SNR G15.4+0.1~\cite{ref:j1818} which was recently discovered at radio wavelengths by
VLA. The VHE $\gamma$-ray emission is extended beyond the H.E.S.S.~PSF ($\sim\!{}6^\prime$), though significantly
less than the shell of the SNR as seen in radio.\\
\par
{\bf Discovery of VHE $\gamma$-ray emission near PSR~J1831-0952}\\
During survey operations, H.E.S.S.~has detected an extended source
(HESS J1831-098)
near the $67\,\mathrm{ms}$ pulsar PSR~J1831-0952~\cite{ref:j1831}. The
source's spectrum is hard with a photon index of $2.1\,\pm\,0.08$ and
with a flux of $\sim 4\,\%$ of the Crab nebula. Adopting the
dispersion measure distance of the pulsar ($4.3\,\mathrm{kpc}$), less
than $1\,\%$ of its spin-down energy would be required to provide the
observed luminosity of the VHE source. The analysis of Fermi data
shows no significant emission above $10\,\mathrm{GeV}$ in coincidence
with the H.E.S.S.~source.\\
\par
{\bf Acknowledgements}\\
\begin{small}
The support of the Namibian authorities and of the University of Namibia
in facilitating the construction and operation of H.E.S.S.~is gratefully
acknowledged, as is the support by the German Ministry for Education and
Research (BMBF), the Max Planck Society, the French Ministry for Research,
the CNRS-IN2P3 and the Astroparticle Interdisciplinary Programme of the
CNRS, the U.K.~Science and Technology Facilities Council (STFC),
the IPNP of the Charles University, the Polish Ministry of Science and 
Higher Education, the South African Department of
Science and Technology and National Research Foundation, and by the
University of Namibia. We appreciate the excellent work of the technical
support staff in Berlin, Durham, Hamburg, Heidelberg, Palaiseau, Paris,
Saclay, and in Namibia in the construction and operation of the
equipment.
\end{small}

\clearpage

\end{document}